\documentclass[conference]{IEEEtran}
\IEEEoverridecommandlockouts
\usepackage{cite}
\usepackage{amsmath,amssymb,amsfonts}
\usepackage{algorithmic}
\usepackage{graphicx}
\usepackage{subcaption}
\usepackage{textcomp}
\usepackage{xcolor}
\usepackage{amsmath, amssymb, bm}
\def\BibTeX{{\rm B\kern-.05em{\sc i\kern-.025em b}\kern-.08em
    T\kern-.1667em\lower.7ex\hbox{E}\kern-.125emX}}
\begin{document}
\bstctlcite{IEEEexample:BSTcontrol}

\title{Quantifying Geometry Effects on Low-Cost Intelligent Reflecting Surfaces}
 \author{\IEEEauthorblockN{Yizhi He$^{1}$, Sayed Amir Hoseini$^{2}$, Mahbub Hassan$^{1}$}\\
 \IEEEauthorblockA{ $^{1}$School of Computer Science and Engineering, University of New South Wales, Sydney, Australia \\
 $^{2}$School of Systems and Computing, University of New South Wales, Canberra, Australia \\
 Email: ethan.he@student.unsw.edu.au,\{s.a.hoseini,mahbub.hassan\}@unsw.edu.au}}

\maketitle

\begin{abstract}
Intelligent Reflecting Surfaces (IRS) promise low-power coverage extension, yet practical deployments must curb hardware complexity and control overhead. This paper quantifies the performance impact of two cost-saving measures, column-wise element grouping and 1-bit (binary) phase quantization, relative to the ideal fully-controlled, continuous-phase baseline. A single-input single-output link is simulated at 26 GHz (mmWave) across three deployment geometries that vary the relative heights of access point, IRS and user equipment. Results show that switching from continuous to binary phase control reduces median SNR gain by approximately 4 dB, while adopting column-wise grouping introduces a similar penalty; combining both constraints incurs approximately 8 dB loss under height-offset deployments. When all nodes share the same height, the degradation from column-wise control becomes negligible, indicating deployment geometry can offset control-granularity limits. Despite the losses, a 32 × 32 column-wise binary IRS still delivers double-digit SNR gains over the no-IRS baseline in most positions, confirming its viability for cost-constrained scenarios. The study provides quantitative guidelines on when simplified IRS architectures can meet link-budget targets and where full element-wise control remains justified.
\end{abstract}


\section{Introduction}
\label{sec:intro}

Wireless networks underpin applications ranging from mobile broadband and the Internet of Things (IoT) to smart cities and autonomous vehicles.  Traffic growth shows no sign of abating: global consumer mobile data volume is forecast to rise by a factor of six to nine between 2023 and 2033~\cite{nokia2023global}.  Among the technologies proposed to sustain this expansion in the future 6G networks, \emph{intelligent reflecting surfaces} (IRS), also called \emph{reconfigurable intelligent surfaces} (RIS), have attracted intense interest for their ability to enhance capacity and coverage with low power and minimal infrastructure~\cite{hu2017potential,rui2021tutorial}.

An IRS is a programmable metasurface whose elements impose controllable phase shifts on incident signals~\cite{attuluri2024design}.  The device is lightweight, energy‐efficient and largely compatible with current network standards~\cite{liu2021reconfigurable}.  After a decade of mainly theoretical work, research momentum has shifted toward prototypes and field trials~\cite{kim2024NSDI,fara2022prototype,dai2020reconfigurable}, bringing practical engineering challenges to the forefront.

\vspace{2mm}\noindent\textbf{Phase resolution constraints.}\;
Most analyses assume each IRS element offers a \emph{continuous} phase range, but present hardware supports only a small set of discrete values~\cite{mao2022group}.  A common low-complexity choice is \emph{1-bit control}, i.e.\ two phase states ($0^{\circ}$, $180^{\circ}$).  Despite its simplicity, a binary‐phase IRS still achieves the asymptotic squared power gain~$\mathcal{O}(N^{2})$ of its ideal counterpart, where $N$ is the number of elements~\cite{rui2020discrete}.

\vspace{2mm}\noindent\textbf{Element grouping constraints.}\;
Large IRS arrays maximise beamforming gain but complicate channel estimation and optimization~\cite{zheng2019intelligent}.  Subarray architectures alleviate this by grouping adjacent elements under a common control signal~\cite{xinyi2023subarray}.  Element grouping trades fine-grained beamforming for lower hardware cost and training overhead~\cite{rui2020grouping}.

\vspace{2mm}\noindent\textbf{Column-wise control in hardware.}\;
Recent prototypes favor \emph{column-wise control}, in which all elements of a column share one reflection coefficient.  Kim~\textit{et\,al.} built NR-Surface, a real-time $16{\times}16$\,mmWave IRS with column-wise, 1-bit control and microwatt power consumption~\cite{kim2024NSDI, li2025mmwave}.  Park~\textit{et\,al.} demonstrated a $100\,$GHz surface using a similar architecture~\cite{Park2025Column}.  These studies show that grouped, binary-phase IRS designs are technically and economically attractive.

\vspace{2mm}\noindent\textbf{Gap in current literature.}\;
The current literature lacks a comprehensive performance evaluation and comparison of fully controlled continuous IRSs and column‑wise binary IRSs under diverse 3D deployment scenarios.


\vspace{2mm}\noindent\textbf{Contributions.}\;
We close this gap by analysing the performance of low-cost IRSs that employ both constraints over realistic deployment scenarios and carrier frequency (26\,GHz mmWave).  Our main contributions are:
\begin{itemize}
  \item an analytical and simulation framework that jointly models column grouping and binary phase control;
  \item quantitative comparison against continuous-phase and element-wise baselines across three geometries, revealing the SNR penalties of each simplification;
  \item design guidelines showing when column-grouped, 1-bit IRSs still meet link-budget targets and when finer control is warranted.
\end{itemize}

The rest of the paper is organized as follows. Related work is reviewed in Section~\ref{sec:related}. Section~\ref{sec:system} introduces the system model and explains the phase shift optimization approach, including how it is adapted for discretization and column-wise control. Section~\ref{sec:simulation} presents the simulation setup and discusses the performance results. Finally, Section~\ref{sec:con} concludes the paper with a summary of key findings.

\begin{figure*}
    \centering
    \begin{subfigure}[t]{0.32\textwidth}
        \centering
        \includegraphics[width=\linewidth]{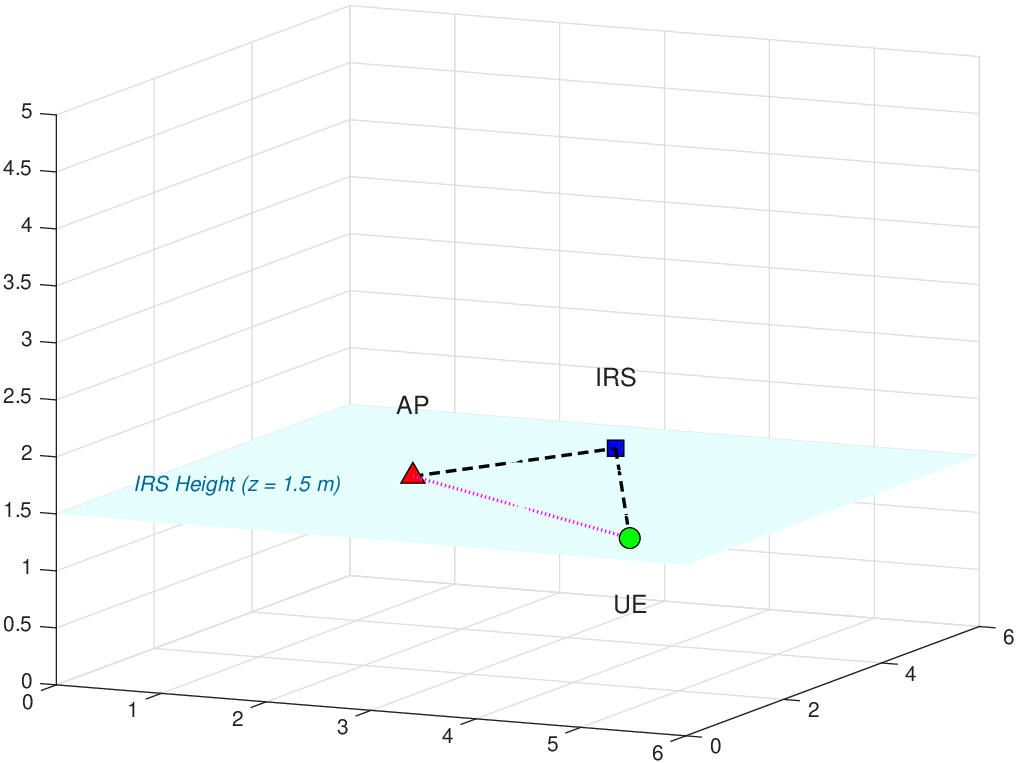}
        \caption{Scenario 1}
        \label{fig:deploy-1}
    \end{subfigure}
    \hfill
    \begin{subfigure}[t]{0.32\textwidth}
        \centering
        \includegraphics[width=\linewidth]{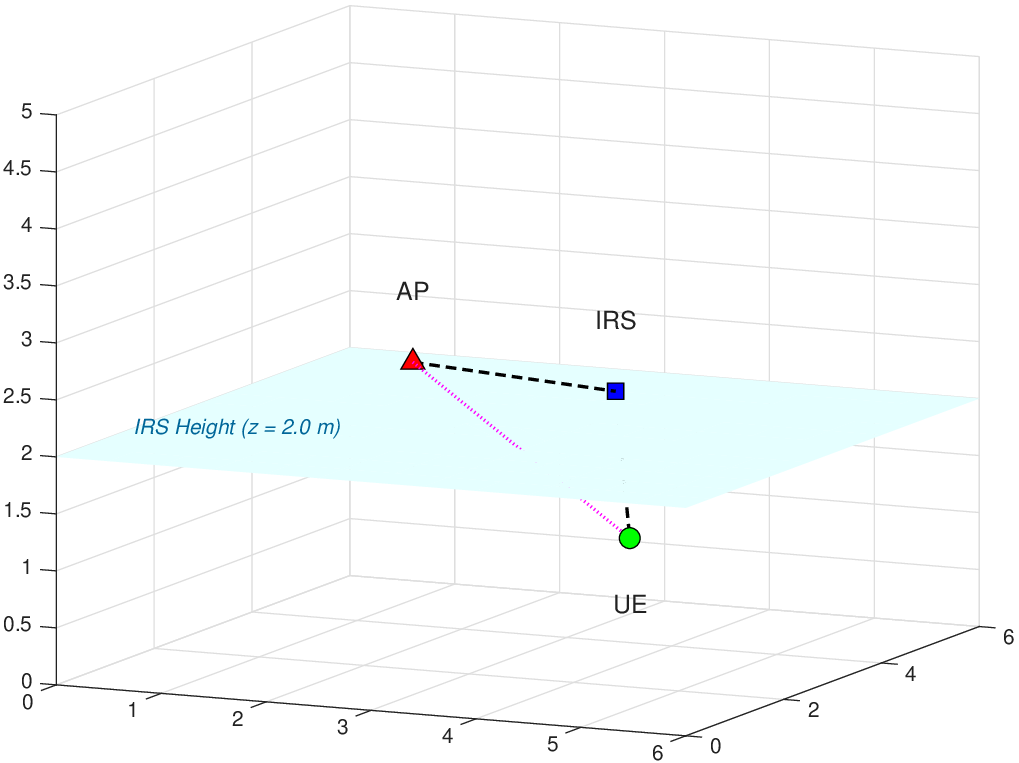}
        \caption{Scenario 2}
        \label{fig:deploy-2}
    \end{subfigure}
    \hfill
    \begin{subfigure}[t]{0.32\textwidth}
        \centering
        \includegraphics[width=\linewidth]{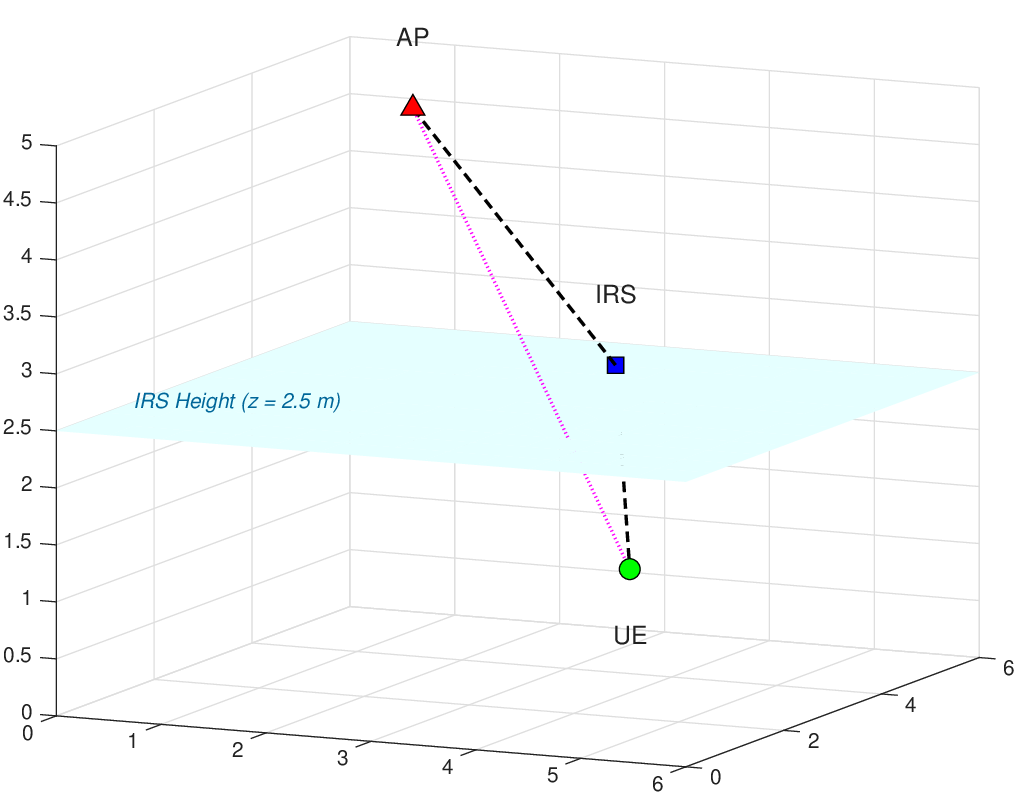}
        \caption{Scenario 3}
        \label{fig:deploy-3}
    \end{subfigure}
    \caption{Three representative scenarios reflecting varied deployments, corresponding to environments such as homes, enterprises, and cellular small cells.}
    \label{fig:deployment-all}
    \vspace{-0.5em}
\end{figure*}

\section{Related Works}
\label{sec:related}
In theoretical analyses, the reflection coefficient of IRS is often modeled with a continuously adjustable phase shift, enabling fine-grained control over signal propagation. However, such idealized assumptions are generally infeasible in practical implementations due to constraints in current hardware technologies and the inherently discrete nature of phase tuning in existing metasurface architectures~\cite{mao2022group}. To address these limitations, discrete phase shift models have been introduced~\cite{rui2020discrete}, which simplify the design and reduce implementation complexity. A prominent example is the 1-bit phase control scheme, wherein each IRS element operates with only two phase states (e.g., $0^\circ$ and $180^\circ$). Despite its simplicity, binary-phase IRS has been shown to achieve the same asymptotic squared power gain of $\mathcal{O}(N^2)$ as its continuous-phase counterpart, where $N$ denotes the number of reflecting elements~\cite{rui2020discrete}. 

In parallel, IRSs are typically composed of a large number of passive reflecting elements to maximize beamforming gain. However, this dense configuration introduces substantial complexity in channel estimation and reflection optimization~\cite{zheng2019intelligent}. To address these challenges, subarray-based architectures have been proposed, wherein adjacent elements are grouped into subarrays that share a common reflection coefficient~\cite{xinyi2023subarray}. This approach reduces hardware cost, simplifies control, and lowers power consumption. Although it entails a loss in beamforming granularity, it offers a favorable trade-off between spectral efficiency and energy efficiency. Moreover, grouping elements with spatially correlated channels, as suggested in~\cite{rui2020grouping}, can further reduce training overhead and computational complexity while preserving most of the passive beamforming gain.

Recent research has explored IRS element grouping strategies that extend beyond conventional rectangular subarrays. Among these, column-wise control, where all elements within a column share a common reflection coefficient, has emerged as a promising approach due to its scalability, reduced control complexity, and hardware efficiency. This architecture has been validated through hardware prototypes, demonstrating its practical viability. For instance, Kim et al.~\cite{kim2024NSDI} introduced NR-Surface, a real-time programmable $16\times16$ mmWave IRS, which employs column-wise control and 1-bit phase quantization. The system supports dynamic beam steering with microwatt-level power consumption, highlighting the feasibility of constrained IRS designs. Similarly, Park et al.~\cite{Park2025Column} proposed a $100\,\text{GHz}$ IRS prototype utilizing column-wise phase control, further reinforcing the growing interest in hardware-aligned, energy-efficient IRS architectures.

Therefore, growing empirical evidence has demonstrated the practicality of low-cost IRS architectures that utilize column-wise control and binary phase shifting, making them increasingly viable for real-world deployment. Despite this growing interest, existing studies have largely focused on ideal continuous-phase IRS models, discrete phase-shift designs with full element-wise control, or subarray-based configurations tailored to specific applications. To date, there has been no comparative trade-off evaluation that considers the impact of geometry and 3D deployment on the performance of fully controlled continuous IRS versus column-wise binary IRS. This paper addresses this gap by analyzing the performance of such simplified IRS designs across a range of realistic deployment scenarios.


\section{System Model}
\label{sec:system}
\subsection{Ideal IRS}
We consider a single-input-single-output (SISO) wireless communication system consisting of a single-antenna Access Point (AP), a passive Intelligent Reflecting Surface (IRS), and a single-antenna User Equipment (UE). The IRS assists AP-to-UE transmission by introducing controllable phase shifts on the incident signals.

We adopt the channel model introduced in \cite{rui2021tutorial}, where the channels of the AP–IRS, IRS–UE, and AP–UE links are denoted by \( \bm{g} \in \mathbb{C}^{N \times 1} \), \( \bm{h}_r^H \in \mathbb{C}^{1 \times N} \), and \( h_d^* \in \mathbb{C} \), respectively. The IRS reflection matrix is defined as \( \bm{\Theta} = \mathrm{diag}\left(\beta e^{j\theta_1}, \dots, \beta e^{j\theta_N} \right) \in \mathbb{C}^{N \times N} \), where \( \beta \in [0, 1] \) is the amplitude attenuation factor and \( \theta_n \in [0, 2\pi) \) is the phase shift applied by the \( n \)-th reflecting element. The received signal at the UE is expressed as
\begin{equation}
y = \left( \bm{h}_r^H \bm{\Theta} \bm{g} + h_d^* \right) \sqrt{P_t} x + z,
\end{equation}
where \( x \) is the transmitted symbol, \( P_t \) is the transmit power, and \( z \) denotes additive white Gaussian noise at the user receiver modeled as CSCG with zero mean and variance \(\sigma^2\). Accordingly, the signal-to-noise ratio (SNR) for user is written as 
\vspace{-0.3em}
\begin{align} \gamma=&\frac {{P_{t}| \boldsymbol {h}^{H}_{r} \mathbf \Theta \boldsymbol {g}+ {h}^{*}_{d}|^{2}}}{\sigma ^{2}} \nonumber\\=&\frac {{P_{t}| \sum _{n=1}^{N}{h}^{*}_{r,n}\beta _{n} e^{\jmath \theta _{n}} {g}_{n}+ {h}^{*}_{d}|^{2}}}{\sigma ^{2}}.\end{align}
For an ideal IRS with element-wise continuous phase shifts, the optimal phase configuration is designed to align the reflected signals constructively with the direct AP-UE path. Based on \cite{rui2021tutorial}, the optimal phase shift for each element is given by:
\begin{equation}
\theta_n^{\star} = \mathrm{mod}\left[\zeta - (\phi_n + \psi_n), 2\pi\right], \quad n = 1, \dots, N,
\label{eq:optimization}
\end{equation}
where \( \phi_n \) and \( \psi_n \) are the phases of the \( n \)-th elements of \( \bm{h}_r^H \) and \( \bm{g} \), respectively, and \( \zeta \) is the phase of \( h_d^* \).

\subsection{Practical Constraints}
In this work, we consider a passive IRS where the reflection amplitude \( \beta \) is not controllable and is typically fixed. For simplicity, we assume unit amplitude reflection (i.e., \( \beta = 1 \)), which models a passive IRS without loss.

To implement the binary phase shift constraint considered in this paper, each IRS element is limited to a phase shift of either \( 0^\circ \) or \( 180^\circ \), corresponding to reflection coefficients of \( +1 \) or \( -1 \), respectively. As a result, the entries in the IRS reflection matrix \( \bm{\Theta} \) are restricted to the binary set \( \{+1, -1\} \). This quantization approach is inspired by the discrete phase shift model proposed in \cite{rui2020discrete}, where the continuous phase shift for each IRS element is projected to the nearest discrete value in a predefined finite set (e.g., \( \{0, \pi\} \) for 1-bit control), by minimizing the Euclidean distance on the complex unit circle. In contrast, we adopt a simplified yet effective method by quantizing the ideal continuous phase solution \( \theta_n^{\text{opt}} \) based on the sign of its real component:
\begin{equation}
r_n^{\text{binary}} =
\begin{cases}
+1, & \text{if } \Re\left(e^{j\theta_n^{\text{opt}}}\right) \geq 0, \\
-1, & \text{otherwise}.
\end{cases}
\label{eq:quantisation}
\end{equation}
This real-part thresholding approximates the nearest-point quantization in the 1-bit case and provides a lightweight solution suitable for binary phase-shift IRS simulation.

We also adopt a column-wise phase control scheme, where all elements in the same vertical subarray share a common phase shift. The IRS consists of \( N = N_x \times N_y \) reflecting elements arranged into \( N_x \) vertical subarrays (columns), each containing \( N_y \) elements. This column-wise design is inspired by the subarray-based approach described in \cite{xinyi2023subarray}, where the optimal phase of each subarray is computed based on the geometric position of its top-left element. While the original formulation employs analytical expressions for alignment in a square-shape subarray, we adopt a straightforward implementation in which the phase shift for each column is directly derived from the first (topmost) element in that column. Combining the two constraints, we implement a computationally inexpensive model of the IRS with column-wise binary phase shift.

\section{Simulation setup and Results}
\label{sec:simulation}
In this section, we evaluate the performance of the passive IRS with column-wise binary phase shifting using MATLAB’s Phased Array System Toolbox™, which implements IRS modeling based on the methodologies in~\cite{rui2021tutorial, elmossallamy2020reconfigurable}. For comparison, ideal IRS with fully controllable continuous phase shift, IRS with fully controllable binary phase shift, and IRS with column-wise continuous phase shift have also been evaluated. As shown in \figurename~\ref{fig:deployment-all}, we simulated three representative scenarios: 1) an environment where AP, IRS and UE are all located at the same height of \(1.5\,\text{m}\); 2) an environment where AP, IRS and UE are located at close heights of \(2.5\,\text{m}\), \(2\,\text{m}\) and \(1.5\,\text{m}\), respectively; 3) an environment where AP, IRS and UE are located at different heights of \(5\,\text{m}\), \(2.5\,\text{m}\) and \(1.5\,\text{m}\), respectively. These scenarios can depict various environments, including homes, enterprises, and cellular small cells.

We considered a \(20\text{m} \times 20\text{m}\) environment and all coordinates are expressed in meter (m). The AP is located at position \((4,15)\) and uses a fixed transmit power of \(P=50\,\text{mW}\). We first simulate a \(32\times32\) IRS located at position \((10,20)\) and it operates at \(26\text{GHz}\) which represents the n258 band used in 5G millimeter-wave communication. Such IRS with half-wavelength spacings occupies approximately \(18.4\text{cm}\times18.4\text{cm}\) in physical size. The noise power at the receiver UE is \(-60\,\text{dBm}\). A path loss exponent of \(\alpha=2\) and a reference distance \(d_0=1\text{m}\) are assumed for the entire map. The x-y coordinates of AP and IRS are fixed while we ran the simulation for each potential location of UE. The z-coordinate of each node is varied according to the scenario described above. The primary performance metric evaluated in our simulations is the SNR gain achieved through IRS deployment, measured relative to a baseline scenario without IRS.

\begin{figure}[t]
    \centering
    \includegraphics[width=\linewidth]{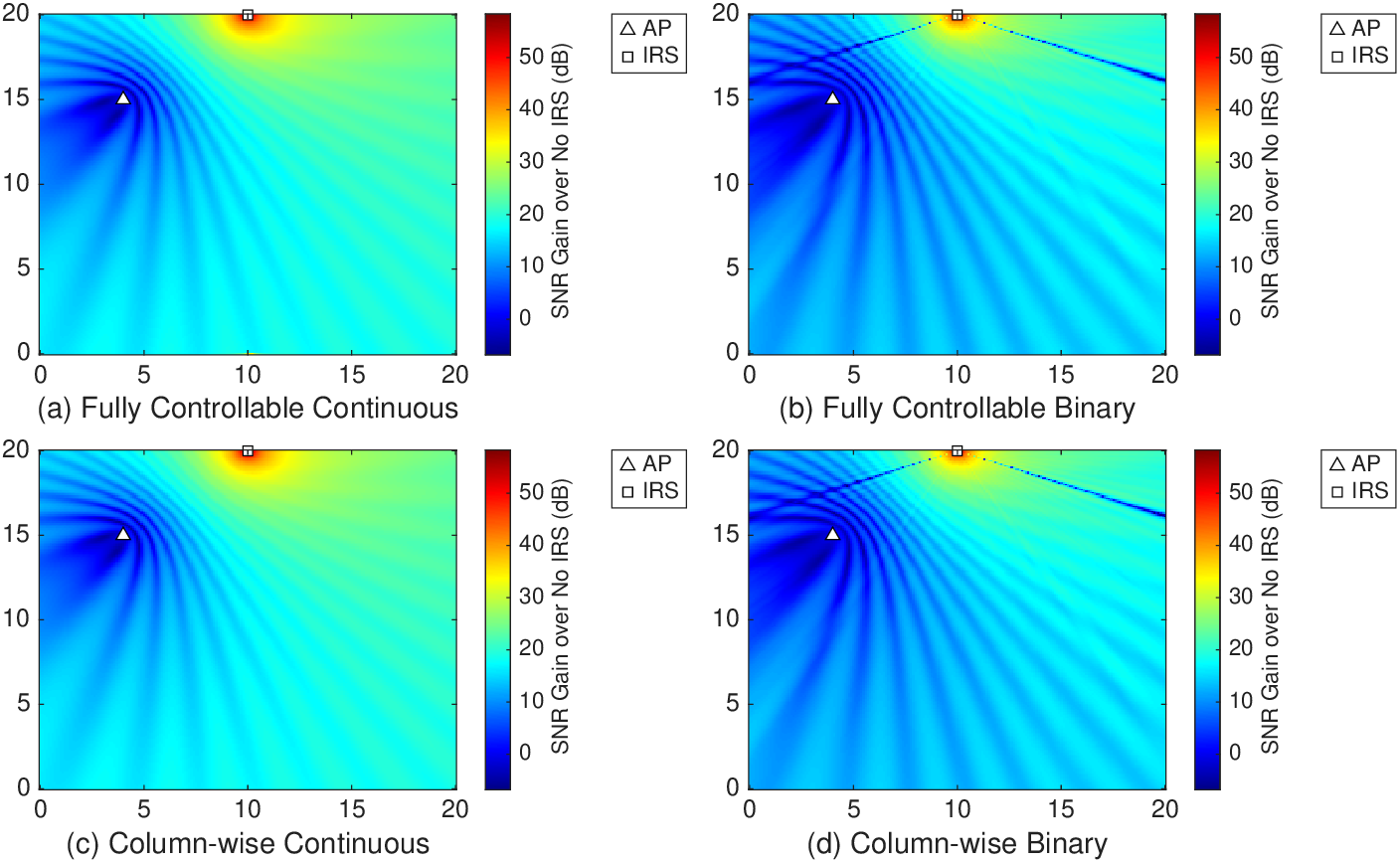}
    \caption{SNR heatmaps for 4 different IRS implementation approaches where AP, IRS and UE share the same height (scenario 1)}
    \label{fig:home-map}
\end{figure}
\begin{figure}[t]
    \centering
    \includegraphics[width=0.9\linewidth]{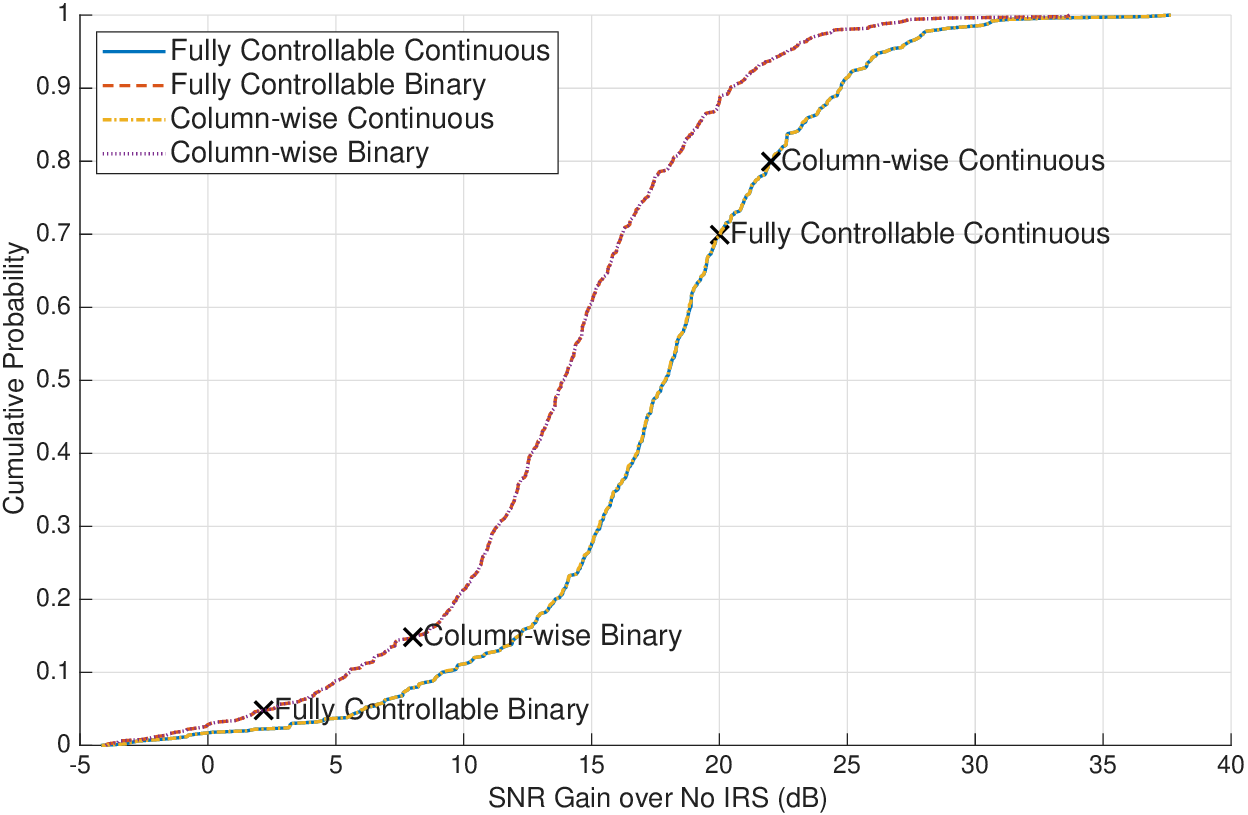}
    \caption{SNR CDF plot for 4 different IRS implementation approaches where AP, IRS and UE share the same height (scenario 1) }
    \label{fig:home-cdf}
    \vspace{-1.5em}
\end{figure}

\figurename~\ref{fig:home-map} shows the SNR gain for scenario 1 for four different implementation approaches as described above.  \figurename~\ref{fig:home-map}a shows identical performance as in \figurename~\ref{fig:home-map}c while \figurename~\ref{fig:home-map}b resembles \figurename~\ref{fig:home-map}d. This illustrates that column-wise configuration does not affect IRS performance when three nodes share the same height. We analyzed the entire map SNR gain statistically and \figurename~\ref{fig:home-cdf} confirm this because the lines of binary phase shift and column-wise binary phase shift are overlapped, while ideal phase shift overlaps with column-wise phase shift. The only factor that downgrades the SNR gain for four IRS implementations is the binary phase shift that drops the SNR gain around 4 dB.

\begin{figure}
    \centering
    \includegraphics[width=\linewidth]{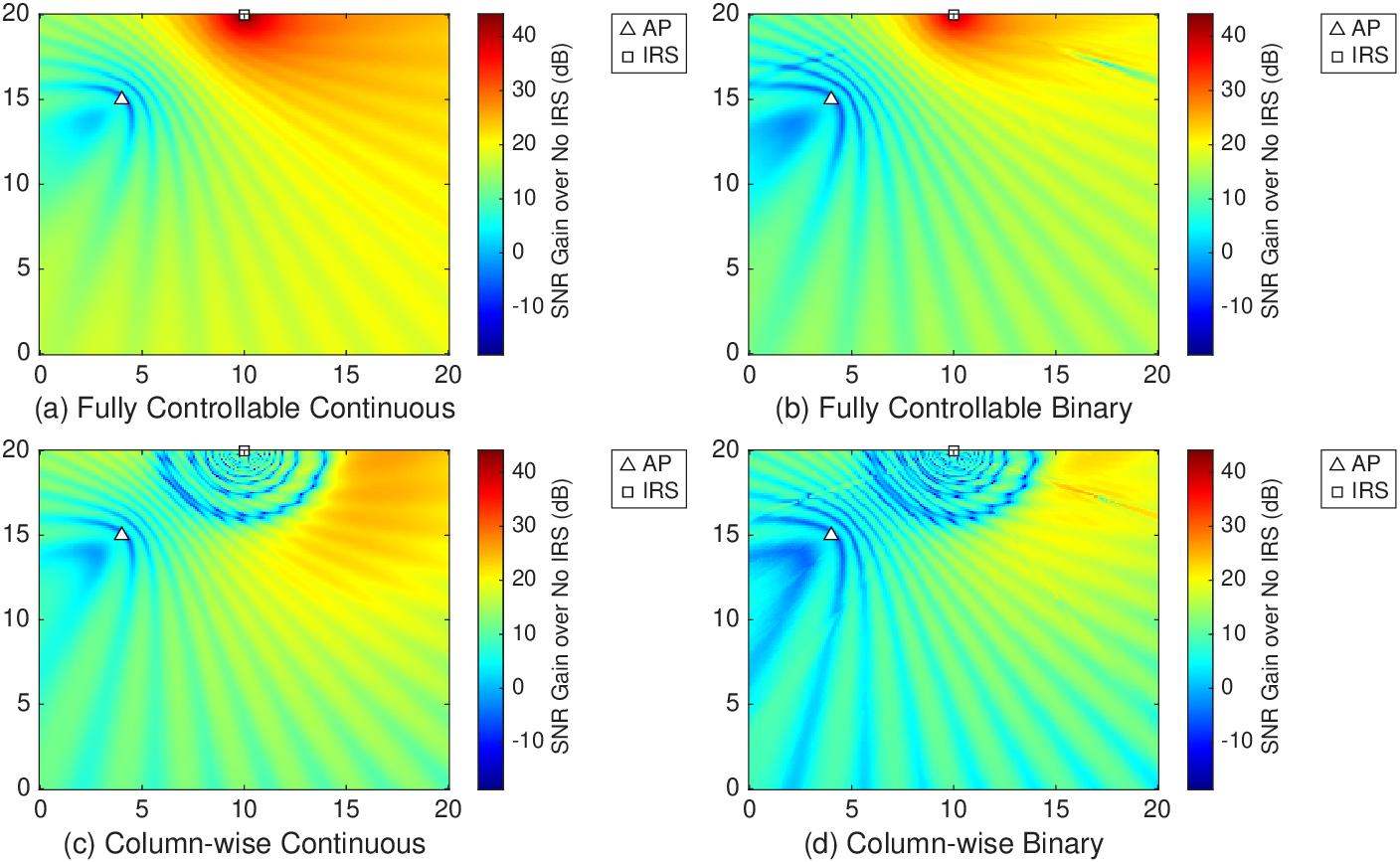}
    \caption{SNR heatmaps for 4 different IRS implementation approaches where AP, IRS and UE are located at heights of 2.5m, 2m and 1.5m. (scenario 2)}
    \label{fig:enterprise-map}
\end{figure}
\begin{figure}[t]
    \centering
    \includegraphics[width=0.9\linewidth]{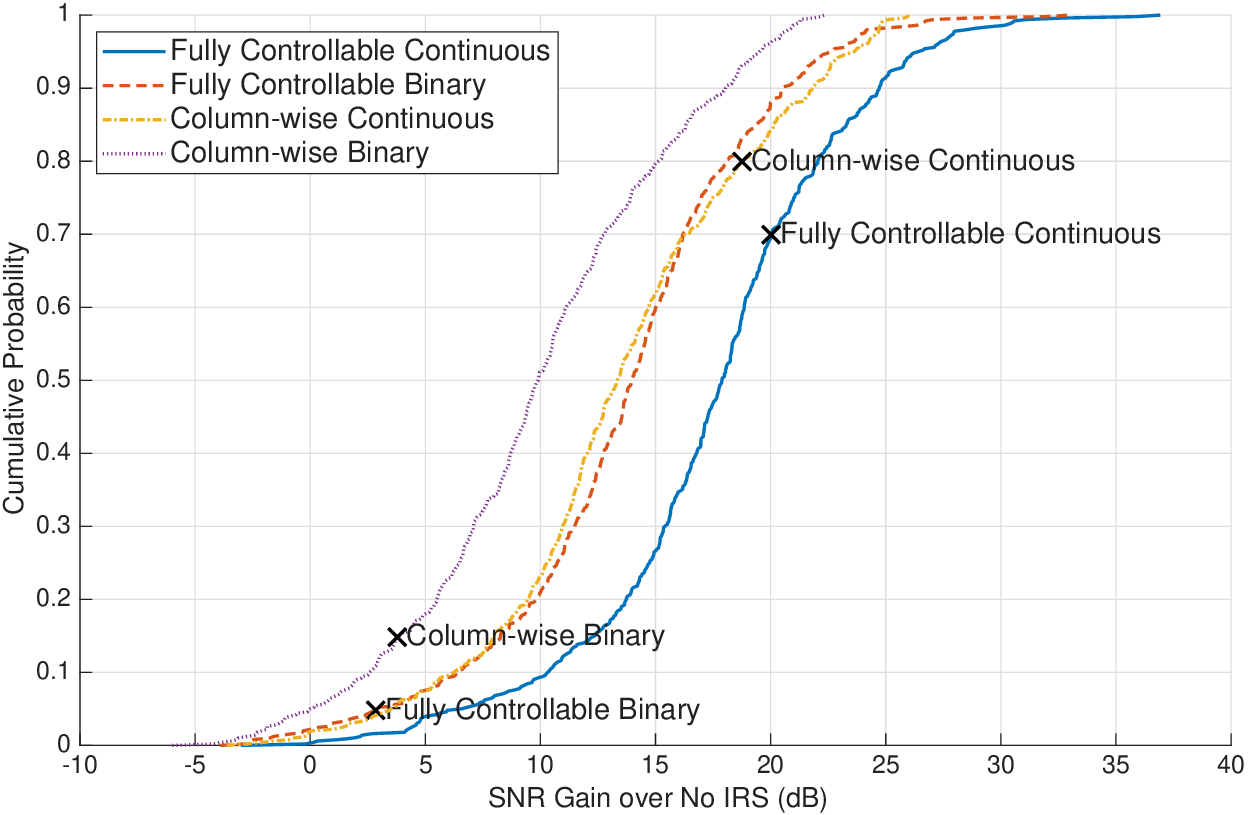}
    \caption{SNR CDF plot for 4 different IRS implementation approaches where AP, IRS and UE are located at heights of 2.5m, 2m and 1.5m. (scenario 2)}
    \label{fig:enterprise-cdf}
    \vspace{-1.5em}
\end{figure}

For scenario 2, where the heights of the nodes are somewhat different, a slight performance degradation is observed by comparing \figurename~\ref{fig:enterprise-map}a and \figurename~\ref{fig:enterprise-map}b, while a more significant degradation is captured between \figurename~\ref{fig:enterprise-map}a and \ref{fig:enterprise-map}c. Also, by looking at \figurename~\ref{fig:enterprise-map}b and \ref{fig:enterprise-map}d, we can observe the SNR gain downgrade between column-wise and non-column-wise IRS. The statistical analysis of scenario 2 in \figurename~\ref{fig:enterprise-cdf} reveals an approximate 4 dB degradation in the SNR gain when transitioning from continuous phase shifting to binary, or from full element-wise control to column-wise control. This degradation increases to around 8 dB when both binary phase shifting and column-wise control are applied simultaneously. 

\begin{figure}
    \centering
    \includegraphics[width=\linewidth]{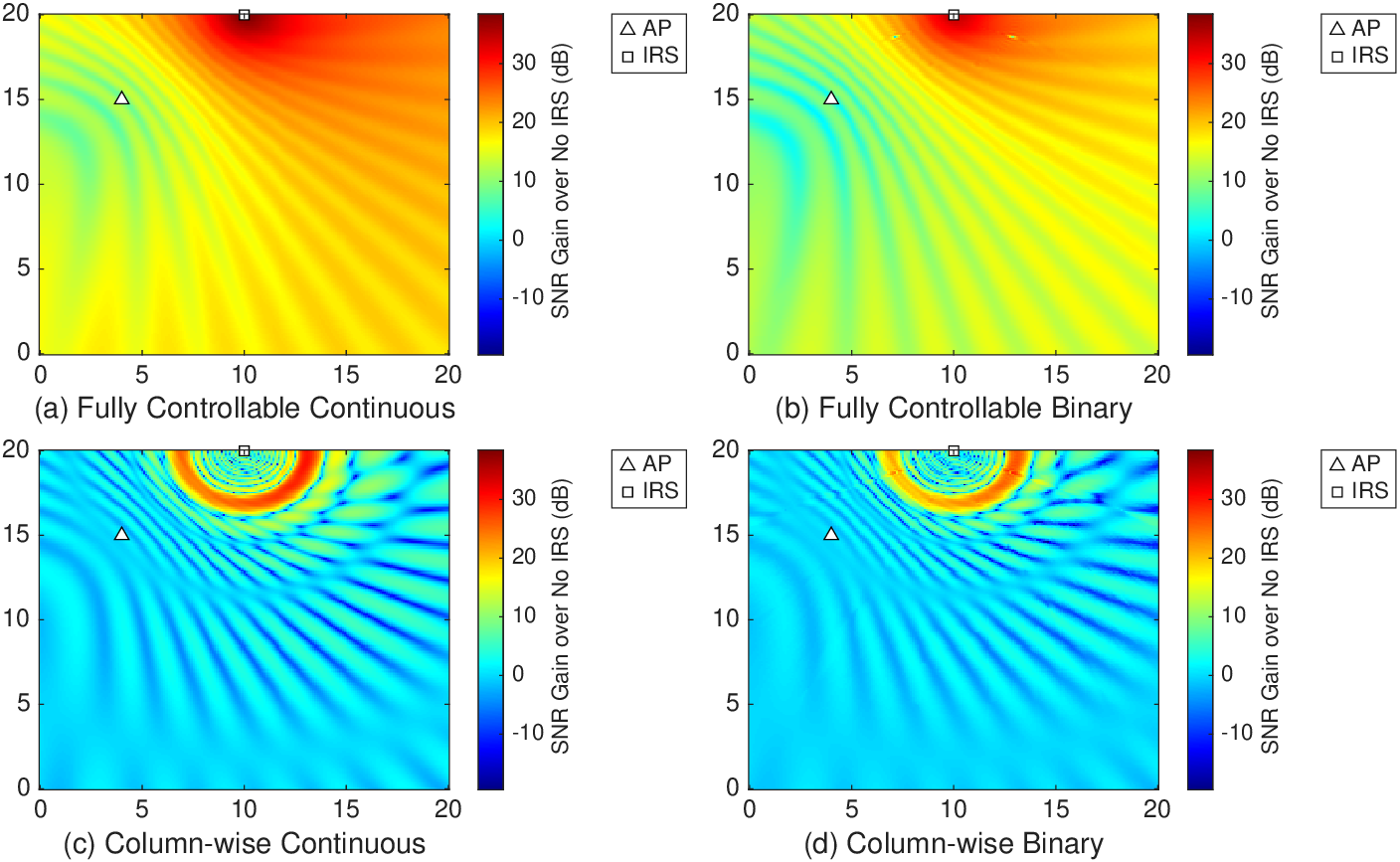}
    \caption{SNR heatmaps for 4 different IRS implementation approaches where AP, IRS and UE are located at heights of 5m, 2.5m and 1.5m. (scenario 3)}
    \label{fig:outdoor-map}
\end{figure}

\begin{figure}
    \centering
    \includegraphics[width=0.9\linewidth]{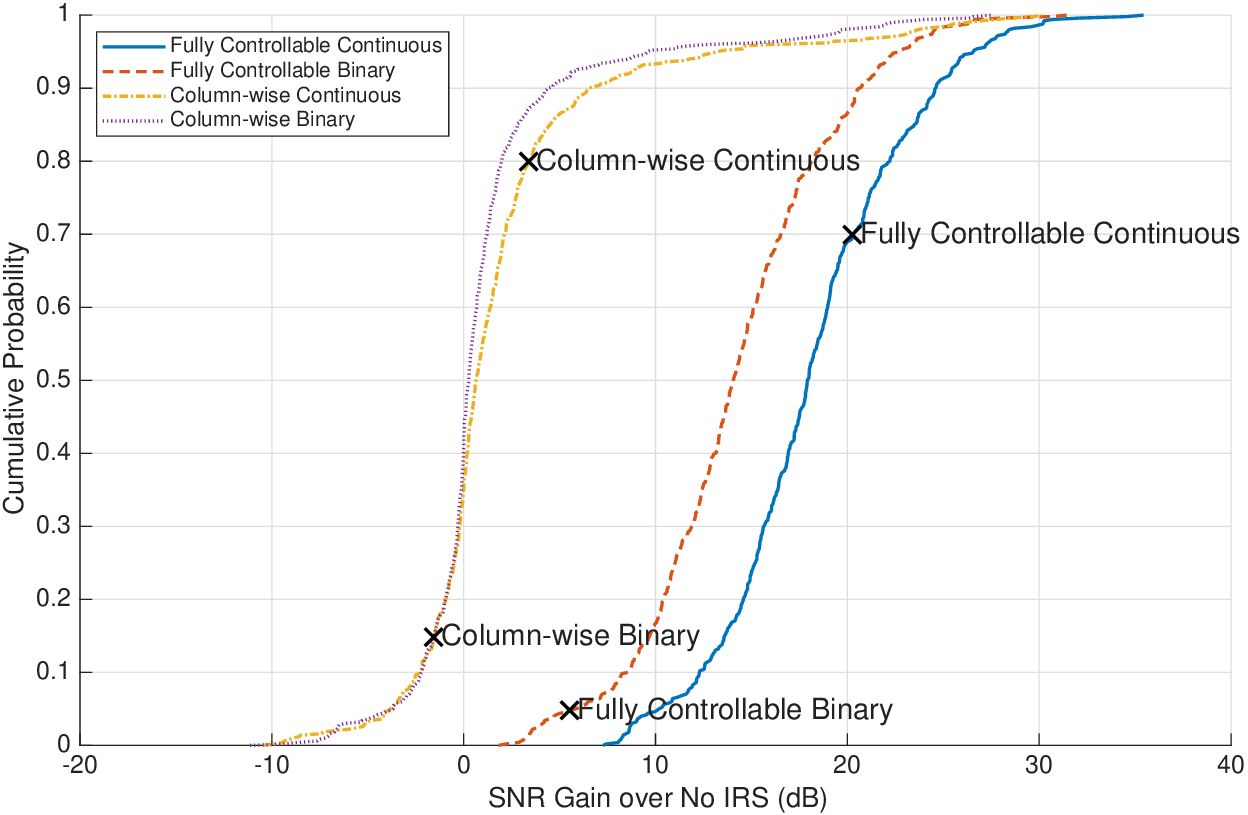}
    \caption{SNR CDF plot for 4 different IRS implementation approaches where AP, IRS and UE are located at heights of 5m, 2.5m and 1.5m. (scenario 3)}
    \label{fig:outdoor-cdf}
    \vspace{-1.1em}
\end{figure}

Scenario 3 represents a more challenging deployment condition, where the AP is installed at an elevated height of 5 meters, typical of a small-cell base station mounted on urban infrastructure such as a streetlight. In \figurename~\ref{fig:outdoor-map}, we can see that 
the non-column-wise configurations are warm colors dominated (i.e., red and orange) as shown in \figurename~\ref{fig:outdoor-map}a and~\ref{fig:outdoor-map}b, whereas the bottom two heatmaps are blue dominant as illustrated in \figurename~\ref{fig:outdoor-map}c and \figurename~\ref{fig:outdoor-map}d that is a significant downgrade in SNR gain for column-wise implementations. 

Also, \figurename~\ref{fig:outdoor-cdf} shows the negative effect of column-wise implementation, where about 40\% of locations in the map do not gain any benefit from the IRS. Furthermore, for roughly 30\% of the area, the column-wise configuration results in a negative SNR gain compared to the no-IRS baseline. The performance gap observed between fully controllable continuous IRS and column-wise binary IRS becomes pronounced in scenario 3. This motivated us to investigate whether the performance degradation is primarily due to the optimization method \eqref{eq:quantisation} we derived for low-cost IRS or the deployment geometry itself. 

\begin{figure}[t]
    \centering
    \includegraphics[width=1\linewidth]{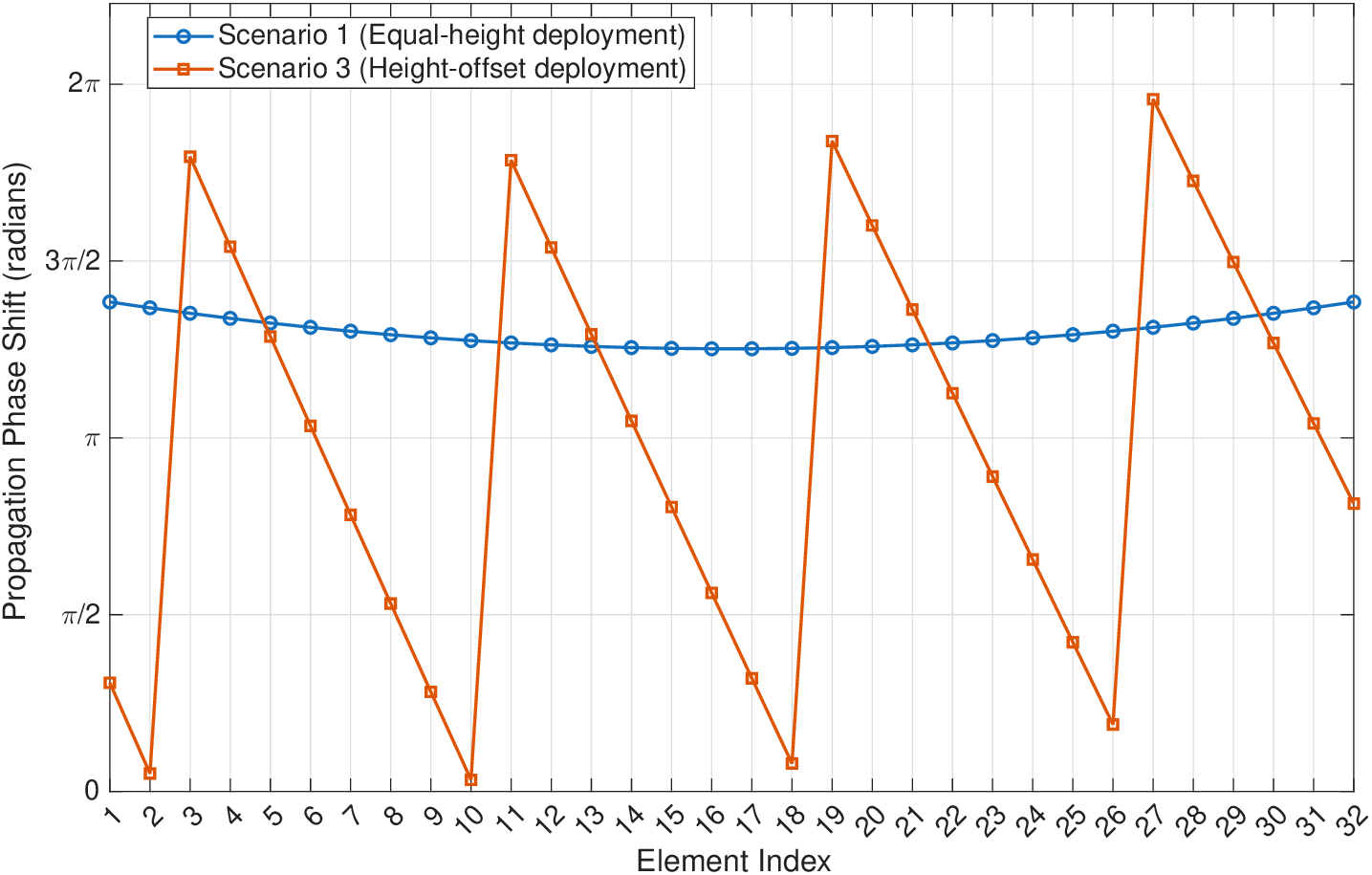}
    \caption{The propagation phase shifts from AP to IRS plus IRS to the UE across a 32-element IRS column under scenario 1 and scenario 3 deployments. }
    \label{fig:phase-lineplot}
\end{figure}

To better understand the performance gap, we analyzed the propagation phase shift of the transmitted signal as it traveled from the AP to each of the 32 elements in a single IRS column, and then from these elements to the UE. As illustrated in \figurename~\ref{fig:phase-lineplot}, we observe that scenario 1 yields an almost flat phase profile across the 32 elements, showing that all elements experience almost identical propagation paths. This implies that assigning a single column-wise coefficient introduces negligible error, which explains why column-wise binary IRS performs close to a fully controllable IRS in this case. In contrast, scenario 3 exhibits a pronounced phase gradient across the column, confirming that different elements experience significantly different propagation delays. Under such conditions, a single column-wise coefficient cannot align all elements simultaneously, leading to noticeable signal misalignment at the UE.

\begin{figure}
    \centering
      \vspace{-1.1em}
    \includegraphics[width=1\linewidth]{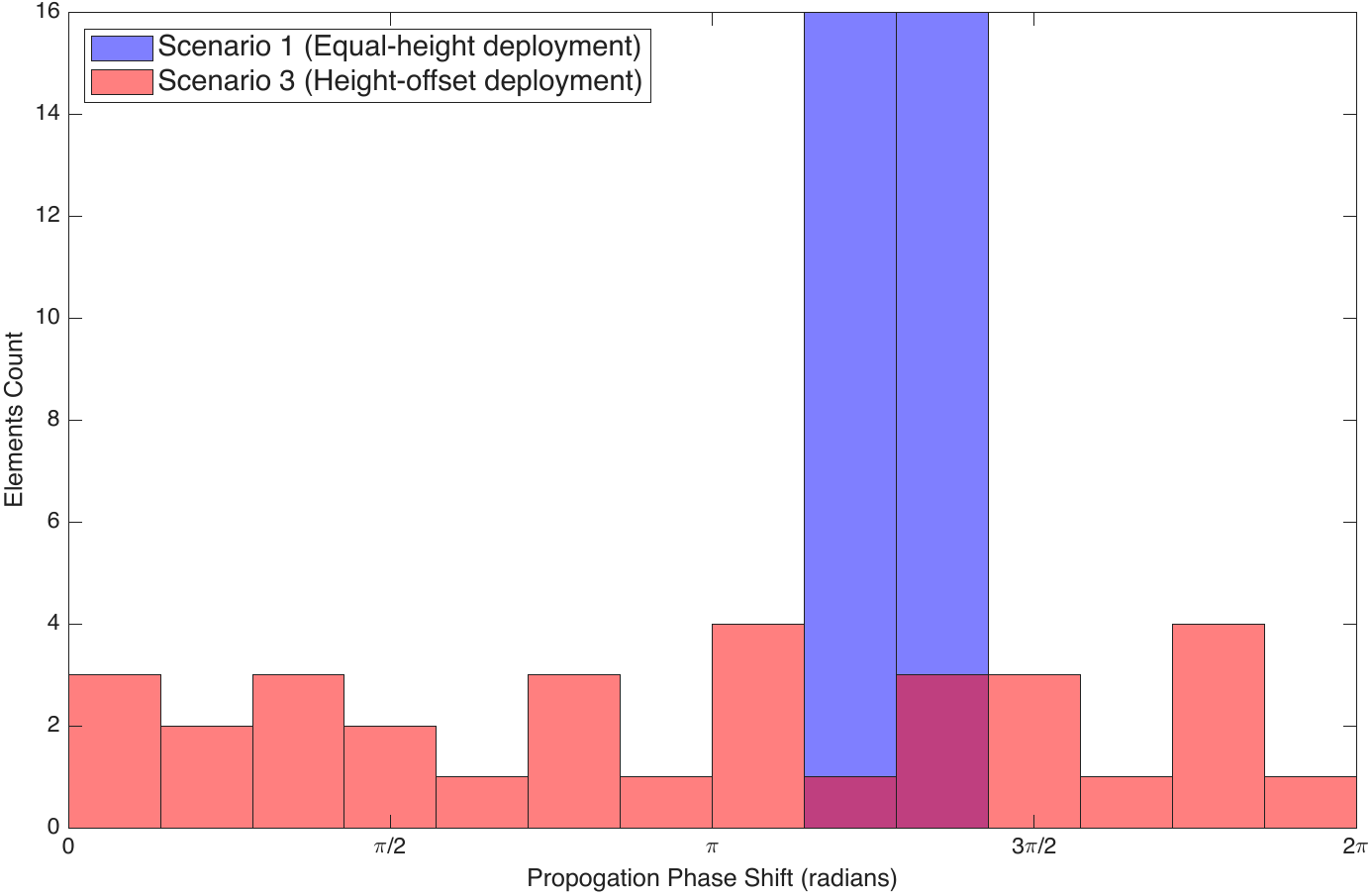}
    \caption{The distribution of the propagation phase shift received at the UE across a 32-element IRS column under scenario 1 and scenario 3 deployments.}
    \label{fig:phase-histogram}
    \vspace{-1.1em}
\end{figure}

\figurename~\ref{fig:phase-histogram} complements this analysis by presenting the distribution of propagation phase shifts for the elements within the same IRS column in the form of a histogram. In scenario 1, the phases cluster tightly around \(\tfrac{5\pi}{4}\), reinforcing that the propagation conditions are effectively coherent across the column. In scenario 3, however, the distribution is broad and dispersed, highlighting that the elements within a column no longer share a common phase reference. This wide spread directly illustrates why column-wise control cannot constructively align the reflected signals at the UE, thereby causing significant SNR degradation.

\begin{figure}
    \centering
    \vspace{0.5em}
    \includegraphics[width=1\linewidth]{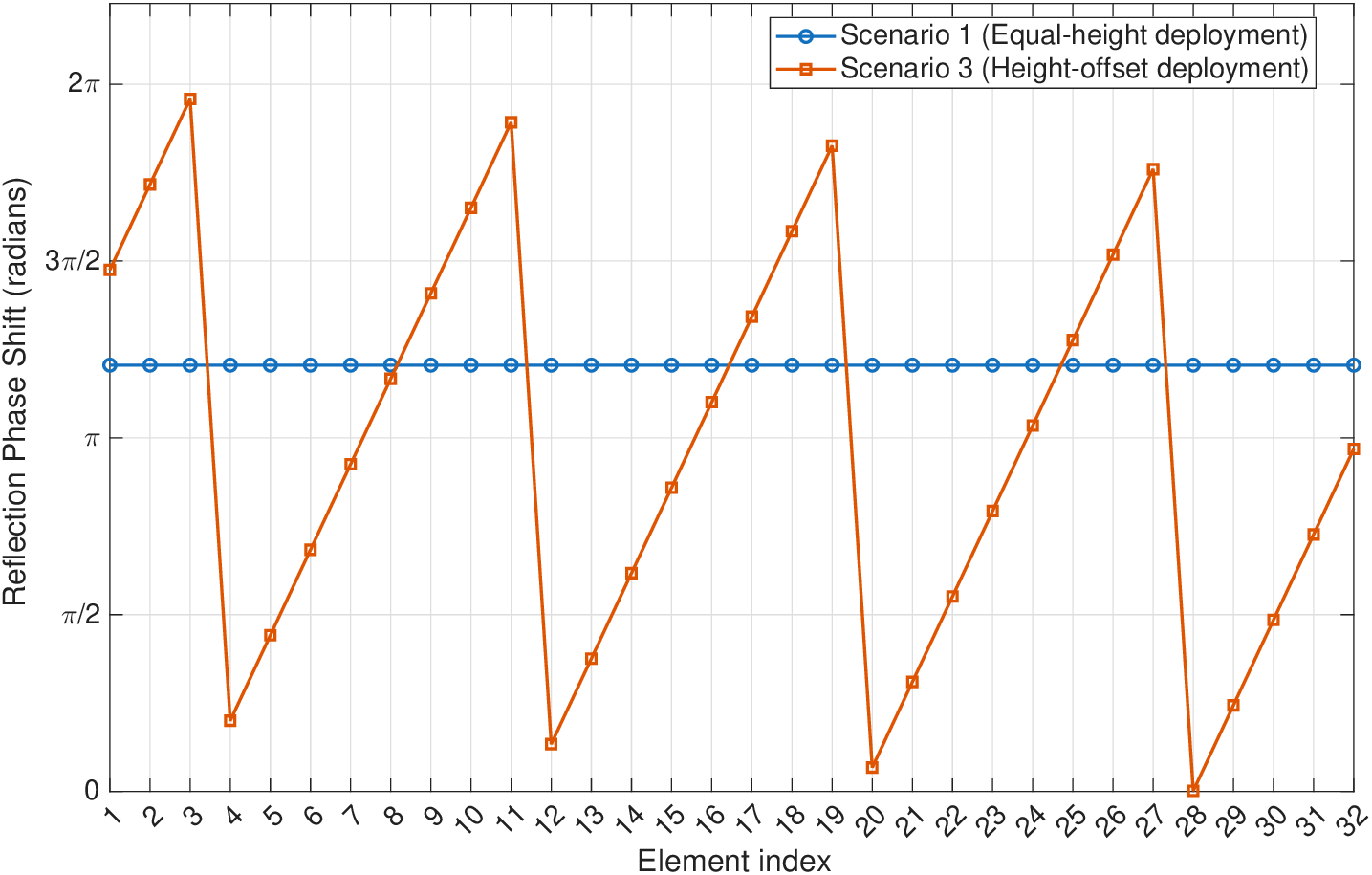}
    \caption{Reflection phase shift applied by a 32-element IRS column under scenario 1 and scenario 3 deployments. }
    \label{fig:reflection-phase}
\end{figure}

We next examined the phase shifts applied by a fully controllable continuous IRS using optimization \eqref{eq:optimization}. \figurename~\ref{fig:reflection-phase} illustrates that, unlike column-wise binary IRS which can only assign one constant coefficient per column, a fully controllable IRS tailors the phase shift at each element to constructively align the LOS signal phase. This ability to compensate for the propagation path differences explains its superior performance in scenario 3.

\begin{figure}
    \centering
    \vspace{0.5em}
    \includegraphics[width=1\linewidth]{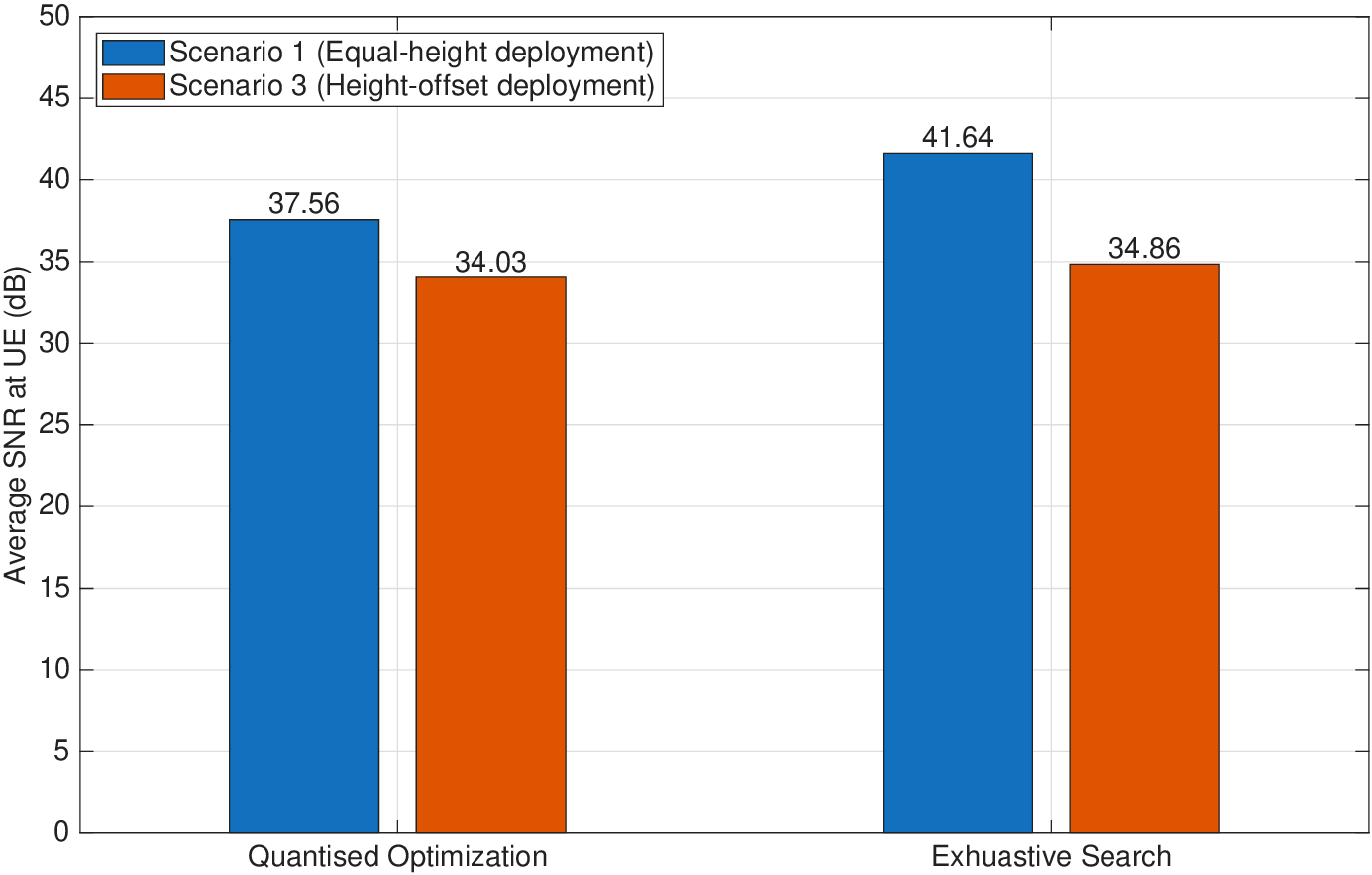}
    \caption{Average SNR across 100 randomly positioned UEs under Scenario~1 (equal-height) and Scenario~3 (height-offset) deployments for column-wise binary IRS with different phase configuration methods.}
    \label{fig:optimization-barplot}
\end{figure}

Finally, \figurename~\ref{fig:optimization-barplot} compares quantized optimization \eqref{eq:quantisation} with exhaustive search for column-wise binary IRS. The exhaustive search evaluates all possible binary phase coefficients to identify the optimal coefficient for each column. The results confirm that the observed performance loss is not caused by errors introduced by quantization of the continuous optimization. Under equal-height deployment, quantization leads to only approximately 4 dB loss compared with exhaustive search. Under height-offset deployment, even exhaustive search fails to close the gap, indicating that deployment geometry, rather than optimization accuracy, is the dominant factor limiting column-wise binary IRS performance.

Simulation results demonstrate that column-wise grouping introduces far greater degradation than binary quantization when the spatial alignment of system components differ significantly. However, when the AP, IRS, and UE are positioned at approximately the same height, the negative effects of column-wise grouping diminish, making low-cost IRS practical in equal-height deployments.

\section{Conclusion}
This work has examined how two common cost-reduction levers, column-wise element grouping and 1-bit phase quantisation, affect the link-budget benefits of Intelligent Reflecting Surfaces. Using a physics-based channel model at 26 GHz across three practical deployment geometries, we bench-marked each hardware simplification against an ideal, fully continuous, element-wise controlled IRS.

The study shows that binary phase resolution alone lowers the median SNR gain by roughly 4 dB, while column-wise control introduces a comparable penalty; applying both constraints together yields an $\sim 8\,$dB reduction under height-offset deployments. Despite this loss, a 32 × 32 column-wise binary IRS still delivers double-digit SNR improvements over the no-IRS baseline in most user positions, confirming its suitability for many cost-sensitive scenarios. A key insight is that height symmetry among access point, IRS and user equipment cancels much of the degradation from column grouping, suggesting that careful placement can compensate for limited control granularity.

These results provide quantitative design guidelines: full element-wise, multi-bit phase control is justified when severe feeder congestion, deep-shadow links or asymmetric mounting heights demand every decibel; otherwise, column-wise binary surfaces offer an attractive, low-power alternative. Future work will extend the analysis to multi-input multi-output (MIMO) links, dynamic user mobility, and real-hardware measurements to verify the simulated trends.

\label{sec:con}


\bibliographystyle{IEEEtran}
\bibliography{ref}

\end{document}